\documentclass[nofootinbib,amsmath,amssymb,amsfonts,aps,prd,eqsecnum,superscriptaddress]{revtex4-1}

\pdfoutput=1
\usepackage{hyperref}
\usepackage{amsmath}
\usepackage{amssymb}
\usepackage{mathrsfs}
\usepackage{xcolor,graphicx}
\usepackage[caption=false]{subfig}
\usepackage[toc,page]{appendix}
\usepackage[section]{placeins}
\usepackage{breqn}
\usepackage[T1]{fontenc}

\newcommand{\vc}{\boldsymbol} 
\newcommand{\la}{\left\langle} 
\newcommand{\ra}{\right\rangle} 

\begin{document}

\title{Numerical Measurements of Scaling Relations in Two-Dimensional Conformal Fluid Turbulence}

\author{John Ryan Westernacher-Schneider}
\email{jwestern@uoguelph.ca}
\affiliation{Department of Physics \\
  University of Guelph \\
  Guelph\, Ontario N1G 2W1\, Canada}
\affiliation{Perimeter Institute for Theoretical Physics \\
  31 Caroline Street North \\
  Waterloo\, Ontario N2L 2Y5\, Canada}

\author{Luis Lehner}
\email{llehner@perimeterinstitute.ca}
\affiliation{Perimeter Institute for Theoretical Physics \\
  31 Caroline Street North \\
  Waterloo\, Ontario N2L 2Y5\, Canada}


\begin{abstract}
We present measurements of relativistic scaling relations in $(2+1)$-dimensional conformal fluid turbulence from direct numerical simulations, in the weakly compressible regime. These relations were analytically derived previously in~\cite{WS2015} for a relativistic fluid; this work is a continuation of that study, providing further analytical insights together
with numerical experiments to test the scaling relations and extract other important features characterizing the
turbulent behavior. We first explicitly demonstrate that the non-relativistic limit of these scaling relations reduce to known results from the statistical theory of incompressible Navier-Stokes turbulence. In simulations of the inverse-cascade range, we find the relevant relativistic scaling relation is satisfied to a high degree of accuracy. We observe that the non-relativistic versions of this scaling relation underperform the relativistic one in both an absolute and relative sense, with a progressive degradation as the rms Mach number increases from $0.14$ to $0.19$. In the direct-cascade range, the two relevant relativistic scaling relations are satisfied with a lower degree of accuracy in a simulation with rms Mach number $0.11$. We elucidate the poorer agreement with further simulations of an incompressible Navier-Stokes fluid. Finally, as has been observed in the incompressible Navier-Stokes case, we show that the energy spectrum in the inverse-cascade of the conformal fluid exhibits $k^{-2}$ scaling rather than the Kolmogorov/Kraichnan expectation of $k^{-5/3}$, and that it is not necessarily associated with compressive effects. We comment on the implications for a recent calculation of the fractal dimension of a turbulent $(3+1)$-dimensional AdS black brane.
\end{abstract}

\maketitle

\section{Introduction}\label{sec:intro}
Relativistic hydrodynamics has become a subject of increased interest
in recent years. Beyond its relevance in astrophysical scenarios 
(e.g.~\cite{2003rnh..book.....W,Gammie:2003rj,Shibata:2005gp,Baumgarte:2010:NRS:2019374,2013rehy.book.....R,Lehner:2014asa}), 
it has become relevant to the description of quark-gluon plasmas (e.g.~\cite{Luzum:2008cw,vanderSchee:2013pia}) and, 
through the fluid-gravity correspondence, it has found its way into the realm of fundamental gravity 
research~\cite{Baier:2007ix,Bhattacharyya:2008jc}. Intriguingly, this correspondence
has revealed that gravity can exhibit turbulent behavior, and studies of its possible consequences are gaining
interesting momentum~\cite{Eling:2010vr,Carrasco:2012nf,Chesler:2013lia, Galtier:2017mve,FIXINGTHEORIES}.
The understanding of turbulence in any regime is a difficult task given its intrinsic complexity, and despite
a long history of efforts in the subject, our knowledge of this rich phenomenon is still incomplete. Important headways into
this subject have been made thanks to statistical analysis complemented with 
numerical simulations (e.g.~\cite{Benzi,Frisch:1996,Dobler:2003,Boffetta:2012,cardy2008non}).
Of particular interest is to understand the possible onset of turbulence, and especially to derive scaling relations in fully-developed scenarios, since those cases are amenable to a statistical description from which
generic statements can be drawn and then tested numerically and observationally.

To date however, only limited attention has been placed on the relativistic turbulent regime, and most of what is
known restricts to the behavior of turbulent, incompressible flows  in the non-relativistic regime.
There has been some work on the analytical front~\cite{Fouxon:2009rd,WS2015,Drivas2017} and several numerical 
investigations~\cite{Carrasco:2012nf,Radice:2012pq,Green:2013zba,WS2015, Chandra:2015iza,MacFadyen2012,MacFadyen2013,East2016,Zhdankin2017, Uzdensky2017}. Because correlation functions can indeed be measured in relevant 
scenarios -- perhaps even in  
QG plasma\cite{Romatschke:2007mq,Luzum:2008cw,Heinz:2011kt,vanderSchee:2013pia,Fukushima:2016xgg} --  and interesting implications for the gravitational field follow from holography, it is of interest to further investigate 
relativistic turbulence. 

In the current work we measure scaling relations in $(2+1)$-dimensional
relativistic conformal fluids in the weakly-compressible turbulent regime and compare them to the predictions in~\cite{WS2015} and various limits thereof. The $(2+1)$-dimensional
case is especially relevant to draw intuition for related phenomena in $(3+1)$-dimensional gravity with the help of the fluid-gravity correspondence (e.g.~\cite{Eling:2010vr,Green:2013zba,FIXINGTHEORIES}). This work is largely a continuation and completion of~\cite{WS2015}, which contained a numerical study which was inconclusive at the time.

This work is organized as follows. Sec.~\eqref{sec:background} provides some background material, discussing both
the inverse- and direct-cascade ranges that could ensue in fully-developed turbulence and, in particular,
relevant scaling relations which we have measured. In Sec.~\eqref{sec:implementation} we provide details of our numerical implementations. This includes the use of a random external force to generate the turbulent flow, as well as considerations specific to simulating either a conformal fluid or an incompressible Navier-Stokes fluid. The equivalence between previously known results and the incompressible limit of the scaling relations derived in~\cite{WS2015} is explicitly demonstrated. We give our results in Sec.~\eqref{sec:results}, where our numerical measurements of the scaling relations derived in~\cite{WS2015} are presented. In Sec.~\eqref{sec:discussion}, we provide
additional discussion and ancillary numerical results, including the demonstration of a $k^{-2}$ energy spectrum in the inverse-cascade of a turbulent conformal fluid which is not due to compressive effects.

Throughout this work, angle brackets $\la . \ra$ will refer to ensemble averages. Letters at the beginning of the alphabet $(a,b,c,\ldots)$ will represent spacetime indices $(0,1,2)$, while letters in the middle of the alphabet $(i,j,k,\ldots)$ will represent spatial indices $(1,2)$. We follow Einstein summation convention. In the context of correlation functions, which often depend on two points $\vc{r}_2$ and $\vc{r}_1$, we define $\vc{r} = \vc{r}_2-\vc{r}_1$. To avoid cumbersome notation, we denote quantities evaluated at $\vc{r}_2$ with a prime (eg. $T_{ij}^\prime$) and quantities evaluated at $\vc{r}_1$ without one (eg. $T_{ij}$). The metric signature is $(-,+,+)$ for our $(2+1)$-dimensional setup. 

%
%
\section{Background}\label{sec:background}
We will make extensive connections with the work presented in~\cite{WS2015}, where
specific scaling relations were derived analytically for $(2+1)$-dimensional relativistic hydrodynamic turbulence. 
We will compare these relations with suitable limits in order to make contact with previously known results, as well as to gauge the importance of relativistic vs compressible contributions in our simulations of the specific case of a conformal fluid.

\subsection{Incompressible non-relativistic limit of the scaling relations}
In this section we explicitly demonstrate that the incompressible Navier-Stokes limit of the relativistic scaling relations 
presented in~\cite{WS2015} can be written in terms of known results. We will use the particular form of a barotropic perfect fluid stress-energy tensor with equation of state $P=\rho/w$, where $P$ is the pressure, $\rho$ is the energy density, and $w$ is the equation of state parameter. In doing so, we obtain incompressible counterparts to the relativistic scaling relations we measure in simulations, which act as a point of reference against which to gauge the relative performance of the relations derived in~\cite{WS2015}.

\subsubsection{Inverse-cascade range} \label{sec:inverseincompressiblelimit}

The first scaling relation, which is valid in the inverse-cascade range, reads~\cite{WS2015}
\begin{eqnarray}
\la T^\prime_{0i} T^{i}_{j} \ra = \frac{1}{2} \epsilon r_j,\label{eq:inversecascade_WS}
\end{eqnarray}
where $\epsilon = \partial_0 \la T_{0i}T_{0}^{i}/2 \ra$. For a $P=\rho/w$ perfect fluid, where $T^{ab} = ((1+w)/w)\rho u^a u^b + (\rho/w) \eta^{ab}$, the scaling relation expands to
\begin{eqnarray}
\la \frac{1+w}{w} \rho^\prime \gamma^{\prime 2} v^\prime_i \left( \frac{1+w}{w}\rho \gamma^2 v^i v_j + \frac{\rho}{w} \delta^i_j \right) \ra = \frac{1}{2} \partial_0 \la \left(\frac{1+w}{w}\right)^2 \rho^2 \gamma^4 \frac{v_i v^i}{2} \ra r_j, \label{eq:inversecascade_WS_expanded}
\end{eqnarray}
where $\gamma$ is the Lorentz factor and $v^i$ is the spatial velocity ($u^a = (-1,v^i)$). In the extreme incompressible non-relativistic limit, $\rho \rightarrow \text{constant}$ and $\gamma \rightarrow 1$. Thus $\rho^\prime = \rho$, and Eq.~\eqref{eq:inversecascade_WS_expanded} becomes
\begin{eqnarray}
\left( \frac{1+w}{w} \right)^2 \rho^2 \la v^\prime_i v^i v_j \ra + \frac{\rho^2}{w} \la v^\prime_i \delta^i_j \ra = \left( \frac{1+w}{w} \right)^2 \rho^2 \frac{1}{2} \epsilon_{\text{NS}} r_j,
\end{eqnarray}
where we have defined $\epsilon_{\text{NS}} \equiv \partial_0 \la v_i v^i \ra/2$ as the incompressible Navier-Stokes version of $\epsilon$. Note that the second term on the left-hand side vanishes due to statistical isotropy, yielding the final result
\begin{eqnarray}
\la v^\prime_i v^i v_j \ra = \frac{1}{2}\epsilon_{NS} r_j .\label{eq:inversecasadelimit}
\end{eqnarray}
Since Eq.~\eqref{eq:inversecasadelimit} is the incompressible Navier-Stokes limit of the relativistic scaling relation in Eq.~\eqref{eq:inversecascade_WS}, they can be compared in the inverse-cascade range of relativistic or compressible turbulence in order to gauge their relative performance.

Notice one can also
arrive at Eq.~\eqref{eq:inversecasadelimit} using known results in the theory of $(2+1)$-dimensional incompressible Navier-Stokes turbulence. In the derivations presented in~\cite{Bernard:1999}, an intermediate result is displayed as
\begin{eqnarray}
\la \delta v_j \delta v_i \delta v^i \ra = 2 \epsilon_{NS} r_j, \label{eq:inversecascade_bernard}
\end{eqnarray}
valid in the inverse-cascade range. Here, $\delta$ denotes a difference, i.e. $\delta v_j = v_j^\prime - v_j$. Expanding out the left-hand side of Eq.~\eqref{eq:inversecascade_bernard} and using statistical homogeneity yields 
\begin{eqnarray}
\la \delta v_j \delta v_i \delta v^i \ra = 4 \la v^\prime_i v^i v_j \ra + 2 \la v_j^\prime v^i v_i \ra.
\end{eqnarray}
By incompressibility, the second term on the right-hand side is divergence-free. Thus, assuming isotropy and regularity at $r=0$, it must vanish~\cite{Landau:1987} (this argument will be used repeatedly in Sec.~\eqref{sec:direct-cascade-limit}). Therefore, Eq.~\eqref{eq:inversecascade_bernard} becomes
\begin{eqnarray}
 \la v^\prime_i v^i v_j \ra = \frac{1}{2} \epsilon_{NS} r_j,
\end{eqnarray}
which is the incompressible non-relativistic limit we obtained in Eq.~\eqref{eq:inversecasadelimit}.

%
%

\subsubsection{Direct-cascade range}\label{sec:direct-cascade-limit}

The second relativistic scaling relation that we consider, which is valid instead in the direct-cascade range, reads~\cite{WS2015}
\begin{eqnarray}
\la \omega^\prime \bar{\omega}_j \ra = -\frac{1}{2}\varepsilon r_j, \label{eq:directcascade_WS}
\end{eqnarray}
where $\omega \equiv \epsilon^{ik} \partial_i T_{0k}$, $\bar{\omega}_j = \epsilon^{ik} \partial_i T_{jk}$, and $\varepsilon \equiv \la \mathcal{F} \omega \ra \equiv \la (\epsilon^{ik} \partial_i f_k) \omega \ra$. Note that $f_k$ is a random external force. In the incompressible non-relativistic limit, $\varepsilon$ becomes proportional to the \emph{enstrophy dissipation rate} $\epsilon_{\omega}$~\cite{Bernard:1999}, namely $\varepsilon \rightarrow ((1+w)/w)^2 \rho^2\epsilon_{\omega}$. Expanding the left-hand side of Eq.~\eqref{eq:directcascade_WS} and setting $\rho=\text{constant}$ and $\gamma=1$ as before, we obtain
\begin{eqnarray}
\la \omega^\prime \bar{\omega}_j \ra &=& \la \epsilon^{ik} \partial_i T_{kj} \epsilon^{mn} \partial^\prime_m T^\prime_{0n} \ra \nonumber\\
&=& \la \epsilon^{ik}\partial_i \left( \frac{1+w}{w}\rho \gamma^{ 2} v_k v_j + \frac{1}{w}\rho \delta_{kj} \right) \epsilon^{mn} \partial^\prime_m \left( \frac{1+w}{w}\rho^\prime \gamma^{\prime 2} v^\prime_n \right) \ra \nonumber\\
&=& \left(\frac{1+w}{w}\rho\right)^2 \la \epsilon^{ik}\partial_i \left( v_k v_j \right) \epsilon^{mn} \partial^\prime_m v^\prime_n \ra + \left(1+w\right) \left( \frac{\rho}{w} \right)^2 \la \delta_{kj} \epsilon^{mn} \partial_m^\prime v_n^\prime \ra.
\end{eqnarray}
The second term on the right-hand side is proportional to the average vorticity, which vanishes by parity invariance. Thus Eq.~\eqref{eq:directcascade_WS} becomes
\begin{eqnarray}
\la \epsilon^{ik}\partial_i \left( v_k v_j \right) \epsilon^{mn} \partial^\prime_m v^\prime_n \ra = -\frac{1}{2}\epsilon_\omega r_j,
\end{eqnarray}
where the non-relativistic vorticity is $\epsilon^{mn} \partial_m v_n \equiv \omega_{\text{NR}}$. The left-hand side needs to be manipulated further in order to compare with standard results (e.g.~\cite{Bernard:1999}). First, notice that the ensemble average on the left-hand side expands under the product rule to
\begin{eqnarray}
\la \epsilon^{ik}\partial_i \left( v_k v_j \right) \epsilon^{mn} \partial^\prime_m v^\prime_n \ra &=& \la \omega_{\text{NR}}^\prime \omega_{\text{NR}} v_j \ra + \la \epsilon^{ik} v_k \partial_i v_j \epsilon^{mn} \partial^\prime_m v^\prime_n \ra.
\end{eqnarray}
We can show that the second term on the right-hand side is zero as follows:
\begin{eqnarray}
\la \epsilon^{ik} v_k \partial_i v_j \epsilon^{mn} \partial^\prime_m v^\prime_n \ra &=& \left( \delta^{im}\delta^{kn}-\delta^{in}\delta^{km} \right) \la v_k \partial_i v_j \partial^\prime_m v^\prime_n \ra \nonumber\\
&=& \la v_k \partial_i v_j \left( \partial^{\prime i} v^{\prime k} - \partial^{\prime k} v^{\prime i} \right) \ra \nonumber\\
&=& \la \left( v_y \partial_x - v_x \partial_y \right) v_j \left(\partial^{\prime x} v^{\prime y} - \partial^{\prime y} v^{\prime x} \right) \ra \nonumber\\
&=& -\la \omega_{\text{NR}}^\prime \left(\vc{v}\times \vc{\nabla}\right) v_j \ra \nonumber\\
&=& -\la \omega_{\text{NR}}^\prime \epsilon^{ik} v_i \partial_k v_j \ra ,
\end{eqnarray}
where we used the identity $\epsilon^{ik} \epsilon^{mn} = \delta^{im}\delta^{kn} - \delta^{in}\delta^{km}$ in the first line. Again, isotropy and regularity at the origin will imply this vanishes, provided it is divergence-free~\cite{Landau:1987}. Thus, we can compute its divergence and show that it vanishes:
\begin{eqnarray}
-\partial^{\prime j} \la \omega_{\text{NR}}^\prime \epsilon^{ik} v_i \partial_k v_j \ra &=& \partial^{j} \la \omega_{\text{NR}}^\prime \epsilon^{ik} v_i \partial_k v_j \ra \nonumber\\
&=& \la \omega_{\text{NR}}^\prime \epsilon^{ik} \partial^j v_i \partial_k v_j \ra \nonumber\\
&=& \la \omega_{\text{NR}}^\prime \left( \partial^x v_x \partial_y v_x - \partial^x v_y \partial_x v_x + \partial^y v_x \partial_y v_y - \partial^y v_y \partial_x v_y \right) \ra \nonumber\\
&=& \la \omega_{\text{NR}}^\prime \left( \partial_y v_x - \partial_x v_y \right) \partial_i v^i \ra \nonumber\\
&=& 0,
\end{eqnarray}
where we used incompressibility in the second and last lines. The relativistic scaling relation Eq.~\eqref{eq:directcascade_WS} thus reduces in the incompressible Navier-Stokes limit to 
\begin{eqnarray}
\la \omega_{\text{NR}}^\prime \omega_{\text{NR}} v_j \ra = -\frac{1}{2}\epsilon_\omega r_j. \label{eq:directcascadeWS_limit}
\end{eqnarray}
As before, this relation is equivalent to an intermediate standard result from~\cite{Bernard:1999}, namely
\begin{eqnarray}
\la \delta v_j \left( \delta \omega \right)^2 \ra = -2\epsilon_\omega r_j. \label{eq:directcascade_bernard}
\end{eqnarray}
To see this, expand the left-hand side and use statistical symmetries to obtain
\begin{eqnarray}
\la \delta v_j \left( \delta \omega \right)^2 \ra &=& 4 \la \omega_{\text{NR}}^\prime \omega_{\text{NR}} v_j \ra + 2\la v_j^\prime \omega_{\text{NR}}^2 \ra,
\end{eqnarray}
and then note that the second term on the right-hand side vanishes by incompressibility, isotropy, and regularity at $r=0$~\cite{Landau:1987}. Thus Eq.~\eqref{eq:directcascade_bernard} is the same as Eq.~\eqref{eq:directcascadeWS_limit}. Since Eq.~\eqref{eq:directcascadeWS_limit} is the incompressible Navier-Stokes limit of Eq.~\eqref{eq:directcascade_WS}, it can be compared in the direct-cascade range of relativistic or compressible turbulence in order to gauge their relative performance.

Finally, we demonstrate that the relativistic correlation derived in~\cite{WS2015}, which reads
\begin{eqnarray}
\la T^\prime_{0T} T_{LT} \ra = \frac{\varepsilon}{24} r^3, \label{eq:directcascadeWS_weird}
\end{eqnarray}
also reduces to a known result in the incompressible non-relativistic limit. Note that the subscripts $(L,T)$ refer to the longitudinal ($\parallel \vc{r}$) and transverse ($\perp \vc{r}$) directions, respectively. Once again, setting $\rho=\text{constant}$ and $\gamma=1$ yields
\begin{eqnarray}
\la v^\prime_T v_L v_T \ra = \frac{\epsilon_\omega}{24} r^3. \label{eq:directcascadeWS_weird_limit}
\end{eqnarray}
Since Eq.~\eqref{eq:directcascadeWS_weird_limit} is the incompressible Navier-Stokes limit of Eq.~\eqref{eq:directcascadeWS_weird}, they can also be compared in the direct-cascade of relativistic or compressible turbulence in order to gauge their relative performance.

Again, Eq.~\eqref{eq:directcascadeWS_weird_limit} can be obtained from standard results in~\cite{Bernard:1999}. The first intermediate result for the direct-cascade range that we use reads
\begin{eqnarray}
\la \delta v_j \delta v_i \delta v^i \ra = \frac{1}{4}\epsilon_\omega x_j r^2. \label{eq:directcascade2_Bernard}
\end{eqnarray}
Using statistical symmetries, the left-hand side expands to $4 \la v^\prime_i v^i v_j \ra + 2 \la v^\prime_j v^i v_i \ra$, and the second term vanishes due to incompressibility, isotropy, and regularity at the origin~\cite{Landau:1987}. Thus, setting $j=L$ in Eq.~\eqref{eq:directcascade2_Bernard}, we obtain
\begin{eqnarray}
\la v^\prime_L v_L v_L \ra + \la v^\prime_T v_L v_T \ra = \frac{1}{16}\epsilon_\omega r^3. \label{eq:firstpiece}
\end{eqnarray}
We can eliminate the first term on the left-hand side using the well-known $1/8$-law, also derived in~\cite{Bernard:1999} and valid in the direct-cascade range, $\la (\delta v_L)^3 \ra = 6 \la v^\prime_L v_L v_L \ra = (1/8) \epsilon_\omega r^3$. This substitution finally yields Eq.~\eqref{eq:directcascadeWS_weird_limit}.
%
%
\section{Implementation}\label{sec:implementation}
As stated, our goal is to explore scaling relations in conformal fluid turbulence. To ensure a clean inertial
regime is established  to compute the appropriate quantities, we include a driving source. Additionally, we 
ensure the numerical methods employed are consistent with the statistical properties of the flow we want to
study. In this section we describe key aspects of our numerical implementation, beginning with general considerations in Sec.~\eqref{sec:implementation-general}. Following this, we present specific considerations for the incompressible and relativistic cases in Secs.~\eqref{sec:implementation-NS} and~\eqref{sec:implementation-rel}, respectively.
%
%
\subsection{General considerations}\label{sec:implementation-general}
\subsubsection{Stochastic Runge-Kutta}\label{sec:SRKII}
In order to implement a random white noise force in a simulation, a special integration algorithm must be used. Based on the work of Honeycutt~\cite{Honeycutt:1992}, we use a second-order Stochastic Runge-Kutta algorithm (SRKII). The Gaussian random force we use, defined later in Eq.~\eqref{eq:Rspacecorr}, is homogeneous, which means the average and variance of the force at every point in space is the same. Thus the prescription described in~\cite{Honeycutt:1992} is applied to each real space point, producing control over the injection rates in an aggregate sense.
%
%
\subsubsection{Pseudorandom number generation}
The random force we employ requires pseudorandom number generation at every time step. For this purpose, we implement the Intel MKL Vector Statistical Library. In particular, we use the Mersenne Twister~\footnote{With BRNG parameter VSL\_BRNG\_MT19937}~\cite{INTEL} and block-splitting for parallel applications~\cite{survivalguideRNG}. We have checked that the energy spectrum $E(k)$ in steady-state is unaffected by the choice of random number generator by comparing the Mersenne Twister (VSL\_BRNG\_MT19937) and the 59-bit multiplicative congruential generator (VSL\_BRNG\_MCG59). We also checked that the output of our code is system-independent~\cite{survivalguideRNG} by running it on two independent clusters.
%
%
\subsubsection{Defining an injection length scale} \label{sec:injectionlength}
In studies of turbulence, the energy/enstrophy injection and scale play a crucial role in establishing and identifying
particularly relevant dynamical ranges. 
One can define an injection length scale associated with the external force in terms of the injection rates of energy and enstrophy as follows. Given Kraichnan-Batchelor~\cite{Kraichnan:1967} scaling of the energy spectrum in the inverse and direct cascades, $E(k) \sim \epsilon_0^{2/3} k^{-5/3}$, $\eta_0^{2/3} k^{-3}$, respectively, one can take the injection scale to be the wavenumber at which $E(k)$ transitions between these two scalings. Thus, set $\epsilon_0^{2/3} k_f^{-5/3} = \eta_0^{2/3} k_f^{-3}$ and solve to find $k_f = \sqrt{\eta_0/\epsilon_0}$. This definition will accurately represent the injection scale up to a numerical factor of order $\sim 1$, so long as the energy spectrum transitions between these two behaviours over a short range of wavenumbers.
%
%
\subsection{Incompressible case}\label{sec:implementation-NS}
\subsubsection{Formulation}
In the {\em incompressible} Navier-Stokes case in $2$D, the entire dynamics is determined by a single pseudo-scalar quantity, the vorticity $\omega = \vc{\nabla} \times \vc{v}$. Thus, it is computationally more efficient to evolve the vorticity equation directly, rather than the components of the velocity. We write the vorticity equation in ``flux-conservative form",
\begin{eqnarray}
\partial_t \omega + \partial_i (v^i \omega) = f_\omega - \nu_4 \partial^4 \omega,
\end{eqnarray}
where $f_\omega$ is the random force defined in the next section, and the dissipative term $- \nu_4 \partial^4 \omega \equiv - \nu_4 \nabla^4 \omega$ on the right-hand side is often referred in the turbulence literature as ``hyperviscosity of order $4$''. Hyperviscosity is frequently used in simulations of an incompressible Navier-Stokes fluid~\cite{Boffetta:2012}, since it limits the range of scales over which dissipation is active (yielding wider inertial ranges for a given grid resolution).
%
%
\subsubsection{Random force and injection rates}\label{sec:force}

The external force appears as $f_{\omega}\equiv \vc{\nabla}\times\vc{f}$, and we wish to construct $f_\omega$ directly
with the appropriate statistical properties. Given a Gaussian random force with a two-point correlation in real space given by
\begin{eqnarray}
\la f_\omega(t,0) f_\omega(t^\prime,r) \ra = g(r) \delta(t-t^\prime), \label{eq:Rspacecorr}
\end{eqnarray}
for some function $g(r)$, the injection rate of enstrophy will be given by $g(0)/2\equiv \eta_0$~\cite{Novikov:1965}, owing to the delta function (i.e. white noise) and to the choice of Gaussian randomness. Ignoring the temporal part of the correlation, we have in Fourier space
\begin{eqnarray}
\la \hat{f}_\omega (\vc{k}) \hat{f}_\omega^* (\vc{k}) \ra = \hat{g} (k), \label{eq:Fspacecorr}
\end{eqnarray}
where reality of the force in real space requires $f_\omega (-\vc{k}) = f^*_\omega(\vc{k})$. 

In order to specify the enstrophy injection rate $\eta_0$, we use a rescaling strategy as follows. First, define two random scalar fields $A(\vc{k})$, $B(\vc{k})$, with zero average $\la A \ra = \la B \ra = 0$ and unit variance $\la A^2 \ra = \la B^2 \ra = 1$ at all wavenumbers, and set $\hat{f}_\omega (\vc{k}) = A(\vc{k}) + i B(\vc{k})$. We first seek an isotropic rescaling $\hat{f}_\omega \rightarrow \tilde{g}(k) \hat{f}_\omega$ that gives the profile of Eq.~\eqref{eq:Fspacecorr} up to a constant factor. Under this rescaling, $A,B \rightarrow \tilde{g}A,\tilde{g}B$, so the zero average is unchanged but the variance transforms to $\la A^2 \ra, \la B^2 \ra \rightarrow \tilde{g}^2\la A^2\ra, \tilde{g}^2\la B^2\ra = \tilde{g}^2$. Thus,
\begin{eqnarray}
\la \hat{f}_\omega (\vc{k}) \hat{f}_\omega^*(\vc{k}) \ra &=& (A+iB)(A-iB) \, , \nonumber\\
&=& A^2 + B^2 \, , \nonumber\\
&\rightarrow & \tilde{g}^2 (A^2+B^2) \, , \nonumber\\
&=& 2\tilde{g}^2 (k) \, . \label{eq:1strescale}
\end{eqnarray}
Thus choosing $\tilde{g} \propto \sqrt{\hat{g}/2}$ gives the desired spatial profile up to a constant factor. To fix the enstrophy injection rate (as $\eta_0 = g(0)/2$), we seek a second rescaling $\hat{f}_\omega \rightarrow R\hat{f}_\omega$ with $R=$ constant determined as follows. As it stands, Eq.~\eqref{eq:1strescale} will produce an enstrophy injection rate given by half of its inverse Fourier transform evaluated at $r=0$, 
\begin{eqnarray}
\tilde{\eta}_0 \equiv \frac{1}{2} FT^{-1}(2\tilde{g}^2(k))\vert_{r=0}.
\end{eqnarray}
Under the second rescaling, Eq.~\eqref{eq:1strescale} becomes $2 R^2 \tilde{g}^2(k)$. Thus the appropriate rescaling is 
\begin{eqnarray}
R=\sqrt{\eta_0 / \tilde{\eta}_0}.
\end{eqnarray}
If one wishes instead to specify the energy injection rate, simply note that for a solenoidal force $\vc{\nabla}\cdot\vc{f}=0$, we have the spatial part of Eq.~\eqref{eq:Rspacecorr} given by
\begin{eqnarray}
\la f_\omega (0) f_\omega (r) \ra &\equiv & \la f_\omega f^\prime_\omega \ra \nonumber\\
&=& \la \epsilon^{ij} \partial_i f_j \epsilon^{mn} \partial^\prime_m f^\prime_n \ra \nonumber\\
&=& \epsilon^{ij}\epsilon^{mn} \partial_i \partial^\prime_m \la f_j f^\prime_n \ra \nonumber\\
&=& (\delta^{im}\delta^{jn} - \delta^{in}\delta^{jm}) \partial_i \partial^\prime_m \la f_j f^\prime_n \ra \nonumber\\
&=& \partial^i \partial^\prime_i \la f^j f^\prime_j \ra \nonumber\\
&=& -\partial^i \partial_i \la f^j f^\prime_j \ra \nonumber\\
&=& -\nabla^2 \la \vc{f} \cdot \vc{f}^\prime \ra .
\end{eqnarray}
So by solving the Poisson equation $\nabla^2 \la \vc{f}(0) \cdot \vc{f}(r) \ra = -g(r)$ one finds the energy injection rate $\epsilon_0$ from the relation $\la \vc{f}(0) \cdot \vc{f}(r) \ra\vert_{r=0} = 2\epsilon_0$. The rescaling factor $R$ can be chosen appropriately in this case. Extracting these a priori injection rates of energy and enstrophy allows one to define an injection length scale as per Sec.~\eqref{sec:injectionlength}.

For our incompressible simulations of the direct-cascade we use a `rectangular' profile, namely $\hat{g}(k) = 1$ in a narrow range of wavenumbers around $k_f$, zero otherwise.
%
%
\subsubsection{Dealiasing}\label{sec:dealiasing}
The Navier-Stokes equation has a quadratic nonlinearity. Thus, two wavenumbers $k_1$, $k_2$ can interact to populate a third wavenumber $k_3=k_1+k_2$. Since we have a finite range of scales resolved in any simulation, $k_3$ could exceed the largest resolved wavenumber, and thus would become represented on the grid as a lower wavenumber $\mathcal{N}-k_3$ (where $\mathcal{N}$ is the grid resolution). In this case, we say $k_3$ has been \emph{aliased}. Prescriptions exist to avoid such aliasing errors. For a quadratically nonlinear term $F\times G$, if we filter out all wavenumber modes with $k>\mathcal{N}/3$ in $F$ and $G$ prior to multiplication, then filter $F\times G$ in the same manner, we will eliminate all aliasing errors. Such a prescription is known as the $2/3$-dealiasing rule, since one retains $2/3$ of the domain in Fourier space. Analogous dealiasing rules exist for higher-order nonlinearities, with less and less of the domain being retained as the order increases. Thus, full dealiasing becomes computationally prohibitive for higher-order nonlinearities, such as for a relativistic fluid flow.
%
%

\subsection{Relativistic conformal fluid case}\label{sec:implementation-rel}
\subsubsection{Formulation}
The system of equations is given by $\nabla_a T^{ab} = f^b$ and the conformal perfect fluid stress-energy tensor $T^{ab} = (3/2)\rho u^a u^b + (1/2)\rho\eta^{ab}$, which uses the conformal equation of state $P=\rho/2$ in $(2+1)$ dimensions. Defining the conservative variables as $(D,S^i) = (T^{00},T^{0i})$, they appear in terms of the primitive variables as
\begin{eqnarray}
(D,S^i) = \left( \frac{3}{2}\rho\gamma^2-\frac{1}{2}\rho, \frac{3}{2}\rho\gamma^2 v^i \right),
\end{eqnarray}
where $v^i$ is the spatial velocity and $\gamma$ is the Lorentz factor. In terms of these variables, the equations of motion appear in flux-conservative form as
\begin{eqnarray}
\partial_t D + \partial_i S^i &=& 0 \label{eq:Deom} \\
\partial_t S^i + \partial_j (S^j v^i + \frac{1}{2}\rho\delta^{ij}) &=& f^i. \label{eq:Seom}
\end{eqnarray}
We use finite differences to discretize the derivatives, with RK4 in space and SRKII (see Sec.~\eqref{sec:SRKII}) in time. The system is damped at short wavelengths using a 4th-order dissipation scheme discussed in Sec.~\eqref{sec:treatlargek}.

\subsubsection{Random force and injection rates}
We choose the Gaussian white-noise force $f^i$ to be divergence-free by deriving it from a stream function $\psi$, $(f_x, f_y) = (\partial_y \psi, -\partial_x \psi)$. Thus, numerically we build $\psi$ directly in the manner described in Sec.~\eqref{sec:force}. For simulations of the inverse-cascade, we choose
\begin{eqnarray}
\la \psi^\prime \psi \ra = \epsilon l_f^2 \exp{(-r^2/2l_f^2)} \delta(t-t^\prime), \label{eq:relforcestatistics}
\end{eqnarray}
where $l_f$ is the characteristic length scale of the correlation, and  $\epsilon = \la T^{0i} f_i \ra$~\cite{WS2015} is a constant. One can verify the equality $\epsilon = \la T^{0i} f_i \ra$ by applying the 2-dimensional Laplacian to Eq.~\eqref{eq:relforcestatistics}, then noting that the spatial part of $\la f_i (\vc{r}) f^i(0) \ra$, written as $F^i_i \equiv \text{tr}F$, is given by $\text{tr}F = -\nabla^2 \la \psi(\vc{r}) \psi(0) \ra$ and $\text{tr}F = 2\la T^{0i}f_i \ra$~\cite{WS2015}. In the weakly compressible regime, $\epsilon$ is approximately the injection rate of $(1/2)\la T^{0i}T^0_i \ra$, whereas in the incompressible regime it fixes the Newtonian kinetic energy injection rate. 

For simulations of the direct-cascade, we instead choose
\begin{eqnarray}
\la \hat{\psi} \hat{\psi}^{*} \ra \propto 
{\begin{cases}
 1 \:\:\:\: k \sim k_f \, , \\
 0 \:\:\:\: \text{otherwise} \, . \end{cases}}\label{eq:spikeprofile}
\end{eqnarray}
%
%

\subsubsection{Dealiasing} \label{sec:treatlargek}
As alluded to in Sec.~\eqref{sec:dealiasing}, in the relativistic case a full dealiasing is computationally prohibitive. Since the computation of the velocity from the conservative hydrodynamic variables, followed by the computation of the flux, amounts to forming a product of up to $5$ fields, there is a quintic nonlinearity. In the weakly-compressible regime, however, a $2/3$-dealiasing rule would likely eliminate a satisfactory amount of aliasing, since the density and Lorentz factor have a small amount of power at all wavenumbers $k\neq 0$. However, in a future study we wish to explore the strongly compressible and ultrarelativistic regimes where a $2/3$-rule would be inadequate. Thus we opt instead to use a 4th-order numerical dissipation scheme to suppress large wavenumber modes (since we want to explore the suitability of alternative dealiasing strategies for that future study) and employ a sufficiently high resolution (so that possibly
spurious effects stay mainly confined at very high frequencies). For a variable $U$, this scheme amounts to including a term $-\nu_{\text{num}} (\partial_x^4 + \partial_y^4) U$ on the right-hand side of its evolution equation, where $\nu_{\text{num}}>0$ is the strength of the dissipation. It is numerically convenient to write this term as $-\kappa (dx^3 \partial_x^4 + dy^3 \partial_y^4)U$ and control the dissipation strength $\kappa$, as its magnitude will be closer to $1$ and the dissipation length scale will move with the resolution~\cite{gustafsson1995time}.
%
%
\section{Results}\label{sec:results}
In all simulations we use periodic boundary conditions with a box size of $L=2\pi$ and resolution of $\mathcal{N}^2=2048^2$, with a variable step size determined by a CFL condition. This resolution has proven quite adequate for studying correlation functions in both the inverse-cascade (eg.~\cite{Boffetta:2000}) and direct-cascade (eg.~\cite{Pasquero2002,Chen2003}) in incompressible fluid turbulence. We find it is also adequate for the weakly compressible regime studied here.

The time scale over which a turbulent flow is presumed to erase knowledge of its initial conditions is the large-eddy turnover time, which has various interpretations in the literature. Borue~\cite{Borue1994} estimates it as $T=2\pi/\omega_{\text{rms}}$, where $\omega_{\text{rms}}$ is the root-mean-squared vorticity. More generally, we have $T=L/U$ where $L$ is the scale of the largest eddies and $U$ is a characteristic speed at that scale. $L$ is estimated as $2\pi/k_i$, where $k_i$ is the infrared ``cutoff" ($\sim$ largest energy-containing scale), and we estimate $U$ as the root-mean-square of the velocity. In our simulations, these time scales will be quoted for reference. 

Averages will be computed over time, or over independent simulations, or both. The adequacy of the sample sizes is gauged via comparison of the average with the statistical error $\sigma/\sqrt{N}$, where $\sigma$ is the sample standard deviation and $N$ is the sample size. For example, a correlation function $f(r)$ will have an ensemble of values for each $r$, and $\sigma(r)$ is computed as the standard deviation of that collection of values.
%
%
\subsection{Inverse-cascade simulations} \label{sec:inversecascaderesults}
We simulate the inverse-cascade of a $(2+1)$-dimensional conformal fluid with an external force described by Eq.~\eqref{eq:relforcestatistics}, and an injection scale $k_f \equiv 2\pi/L_f \sim 203$ defined by $k_f=\sqrt{\eta_0/\epsilon_0}$, as in Sec.~\eqref{sec:injectionlength}. We consider three cases with the numerical dissipation strength given by $\kappa = (0.05,0.03,0.02)$ (so as to compare results among them) and when quoting properties of each case we will present them in this order. Since the force is somewhat broadband, it has power in the dissipation range of scales. Thus, decreasing the dissipation strength is enough to increase the energy growth rate, and thus the rms Mach number of the flow, $v_{rms}/c_s$, where $c_s$ is the sound speed ($1/\sqrt{2}$ of the speed of light, in our case). Statistical quantities are averaged over ensembles of independent simulations, as well as averaged over an interval of time after the energy passes $k=10$ and before it reaches the box size. Table~\eqref{table:icparams} contains various parameters of the flows, as well as the sample sizes for the joint average over an ensemble and over time.

\begin{figure}[h!]
\centering
\hbox{\hspace{0.8cm}\includegraphics[width=0.9\textwidth]{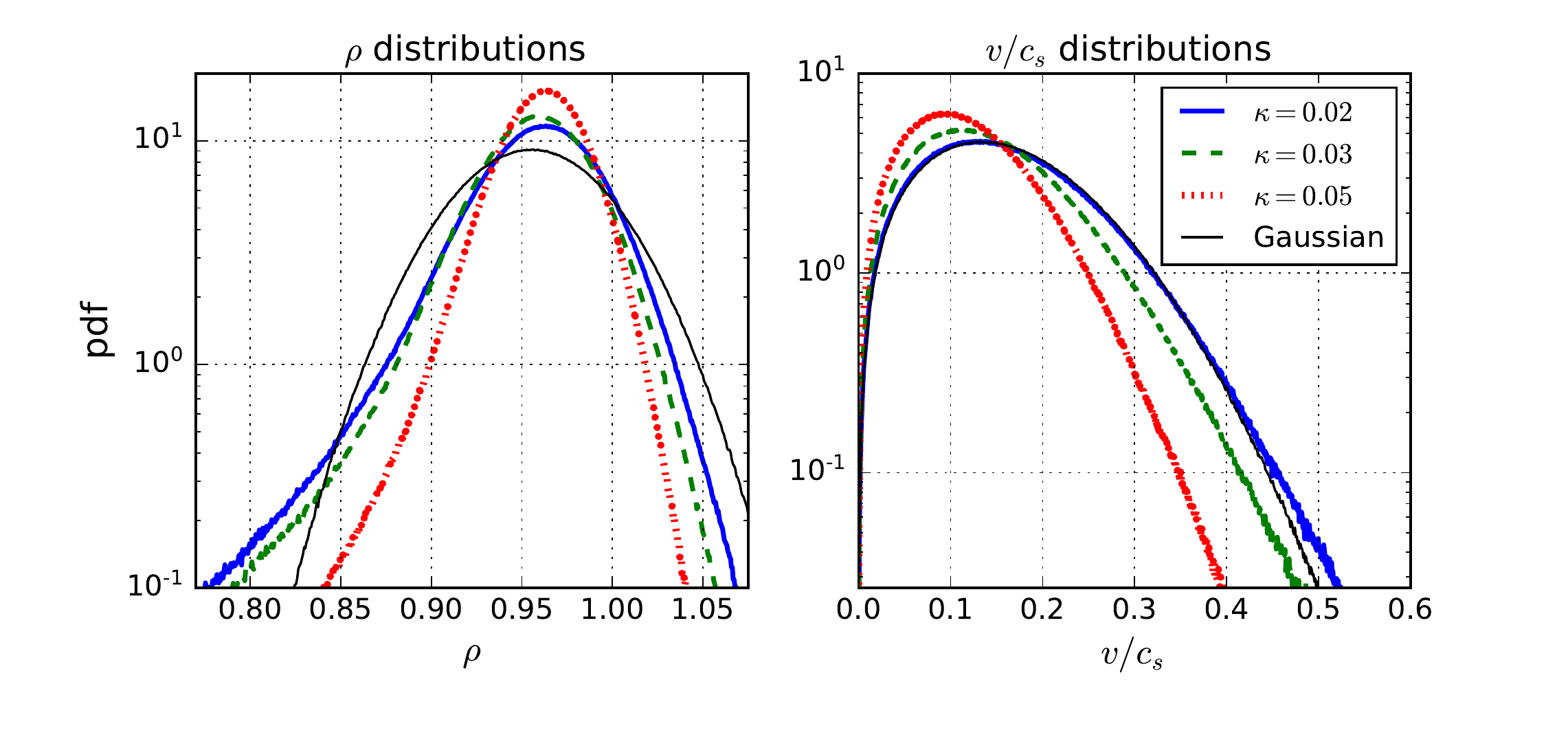}}
\caption{Probability distribution functions for the inverse-cascade simulations. The pdfs of the energy density $\rho$ (Left) and the Mach number $v/c_s$ (Right) are displayed, where $c_s = c/\sqrt{2}$, plotted on a semi-log scale. All dissipation cases are overlaid for ease of comparison. The density $\rho$ and velocity field $(v_x,v_y)$ are high-pass filtered ($k>10$) for a more sensible comparison in this quasi-steady regime (i.e. no large-scale dissipation). The cutoff $k=10$ is chosen based on the maxima of the spectra in Fig.~\eqref{fig:spectra_inverse_cascade} occurring at $k<10$. For comparison, purely Gaussian distributions are plotted (black, solid, thinner lines) with its average and standard deviation matched to data from the dissipation case $\kappa=0.02$. In the order of increasing energy growth rate, the standard deviations of pdf($\rho$) and pdf($v/c_s$) in each case are $(0.0286,0.0395,0.0438)$ and $(0.122,0.147,0.167)$, respectively. In the same order, the rms Mach numbers are $(0.1386,0.1674,0.1893)$. These properties indicate a weakly compressible flow.} \label{fig:pdfs_inverse_cascade}
\end{figure}

In Fig.~\eqref{fig:pdfs_inverse_cascade} we characterize the flows by presenting the probability distributions
functions (pdfs) of the energy density and Mach number. The pdfs are observed to widen as the energy growth rate increases, as one would expect. For comparison, in both cases we also plot Gaussian distributions (black, dashed) with average and standard deviation matched to the data from the $\kappa = 0.02$ case. The Gaussian provides a good fit to the Mach number pdf (although with a slight hint of non-Gaussianity in the tail), whereas the energy density pdf exhibits a stronger, exponential tail towards smaller values.

\begin{figure}[h!]
\centering
\hbox{\hspace{0.8cm}\includegraphics[width=0.9\textwidth]{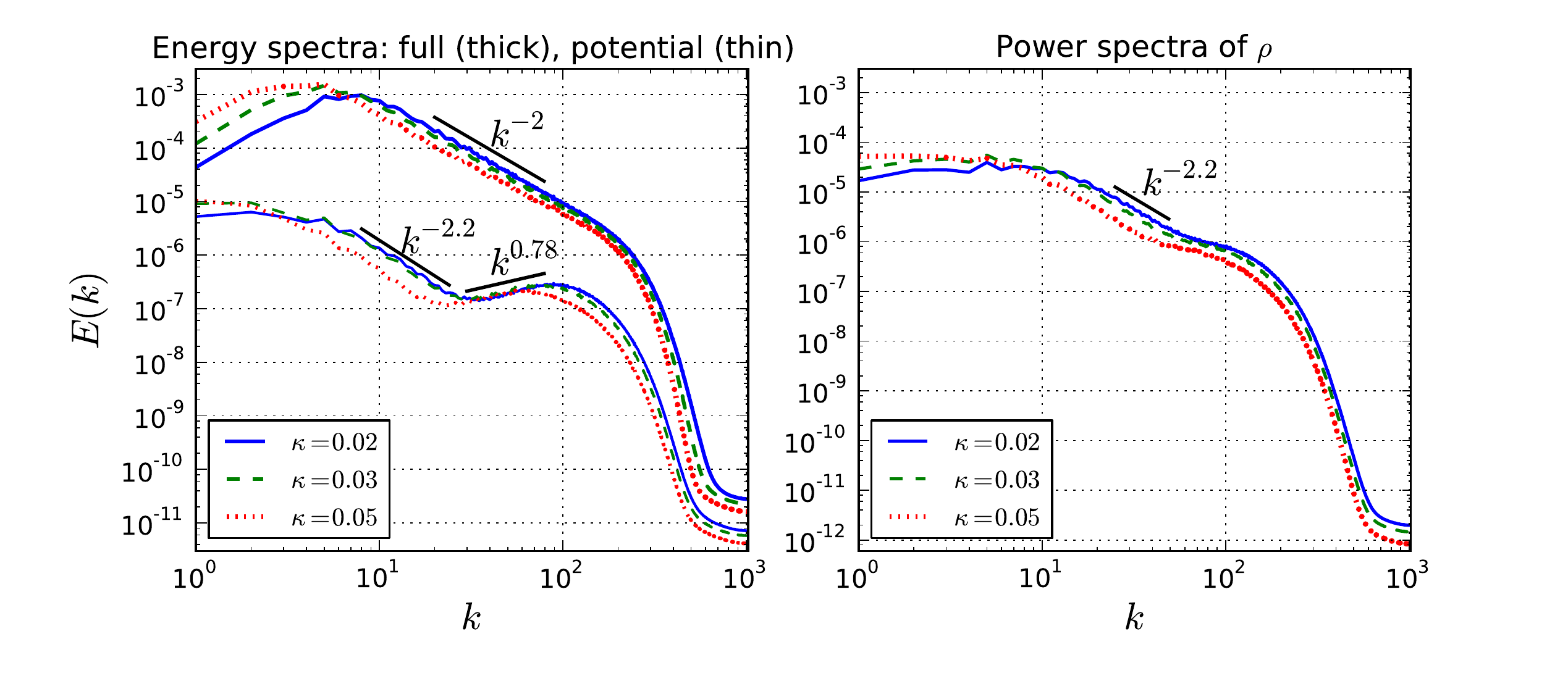}}
\caption{(Left): Newtonian energy spectra of the inverse-cascade simulations plotted on a log-log scale. The spectra corresponding to the full velocity field (thicker lines) and the curl-free potential part (thinner lines) are displayed. All energy growth rate cases are displayed, with the same colour coding and line styles as in Fig.~\eqref{fig:pdfs_inverse_cascade}. The best-fit power laws for the full spectra over the range $k=20-80$ in order of decreasing dissipation strength are $(-1.80,-1.89,-1.96)$. (Right): Power spectra of the energy density $\rho$ for all energy growth rates, plotted on a log-log scale.} \label{fig:spectra_inverse_cascade}
\end{figure}

In Fig.~\eqref{fig:spectra_inverse_cascade} (Left) we display the angle-averaged Newtonian kinetic energy spectra $E(k) \equiv \pi \la \hat{v}^2(\vc{k}) \ra$ (both the full spectrum and the potential part, obtained by projecting the velocity onto $\hat{\vc{k}}$ in Fourier space). We observe a steepening of the inertial range scaling towards $E(k) \sim k^{-2}$, which we note is steeper than the Kolmogorov/Kraichnan power law of $k^{-5/3}$. The spectra are not changed significantly ($<1\%$) by instead using density-weighted velocities $\rho^{1/3} \vc{v}$ or $\rho\vc{v}$, the former having been suggested in the $(3+1)$-dimensional context in~\cite{Kritsuk:2007} to restore Kolmogorov/Kraichnan scaling from the observed spectral exponent of $k^{-2}$. The spectrum of the potential component of the velocity exhibits a bump beginning at $k\sim 30$, with scaling of $k^{-2.2}$ and $k^{0.78}$ on either side. Such a bump towards large $k$ is commonly observed in spectra in simulated compressible flows in $(3+1)$ dimensions, eg.~\cite{Radice:2012pq,Kritsuk:2007,Dobler:2003,Federrath:2013}, and is attributed in those cases to an artefact of high-order numerical dissipation known as the bottleneck effect~\cite{Falkovich:1994}. This effect has also been observed in simulated compressible 2D flows which exhibit transfer of energy to small scales~\cite{Biskamp:1998}. The bump we observe in Fig.~\eqref{fig:spectra_inverse_cascade} is likely due to the same effect, although we cannot make a conclusive statement since we have not performed the specific resolution studies necessary to do so, nor have we used dissipation of a different order. The late-time spectra obtained from our simulations of the direct-cascade (not shown) also exhibit such a bump, and in that case we note that reducing the time step by half does not change the bump perceptibly. 

With regard to the full inverse-cascade spectra in Fig.~\eqref{fig:spectra_inverse_cascade} (Left), it is worth noting that there is no large-scale friction. In~\cite{Scott:2007}, it was shown that the presence of large-scale friction can affect the inertial range spectrum in the incompressible Navier-Stokes case. In the same study it was also shown that measurements of the inertial range spectrum are not reliable without a sufficiently resolved enstrophy cascade ($k_{\text{max}}/k_f \sim 16$, where $k_{\text{max}}$ is defined as $\mathcal{N}/3$). We do not have the direct-cascade range resolved to this degree in Fig.~\eqref{fig:spectra_inverse_cascade} ($k_{\text{max}}/k_f \sim 3.4$). The approach of the full spectrum towards $k^{-2}$ is generally expected for compressible turbulence in both $(3+1)$ dimensions (see eg.~\cite{Federrath:2013}) and $(2+1)$ dimensions (see eg.~\cite{Passot1995}), although usually for much larger Mach numbers than our current simulations. With that said, $(2+1)$-dimensional conformal fluids are special (eg. having a very large sound speed and no mass density), and its turbulent regime is seldom studied (see eg.~\cite{Carrasco:2012nf,Green:2013zba}), so one may not expect the same energy spectra a priori. We elaborate more on this in Sec.~\eqref{sec:discussion}, where we demonstrate that the $k^{-2}$ spectrum is not necessarily associated with compressive effects.

\begin{table}[]
\centering
\caption{Parameters of inverse-cascade simulations: $\kappa$ is the dissipation parameter described in Sec.~\eqref{sec:treatlargek}; $\epsilon$ is the growth rate of $(1/2)\la T_{0i}T^{i}_0\ra$; $v_{rms}/c_s$ is the rms Mach number; $N_t$ is the number of snapshots averaged over time; $N_{ens}$ is the number of independent runs (ensemble size); $2\pi/\omega_{rms}$ is the eddy turnover time defined by the rms vorticity; $L/v_{rms}$ is the eddy turnover time defined by $v_{rms}$ and $L=2\pi/10$; $\delta T$ is the time interval between snapshots of the flow that are averaged over; $T_1$ and $T_2$ are respectively the first the last times over which the temporal average is computed. For comparison, note that the light-crossing time is $2\pi$.}
\label{table:icparams}
\begin{tabular}{llllllllll}
\hline
$\kappa$ & $\epsilon\times 10^{4}$ & $v_{rms}/c_s$ & $N_t$ & $N_{ens}$ & $2\pi/\omega_{rms}$ & $L/v_{rms}$ & $\delta T$ & $T_1$ & $T_2$ \\ \hline
0.05 & 3.3 & 0.1386        & 7     & 20        & 1.15                & 6.411       & 5          & 80    & 110   \\ \hline
0.03 & 5.3 & 0.1674        & 5     & 60        & 0.99                & 5.308       & 5          & 60    & 80    \\ \hline
0.02 & 7.0 & 0.1893        & 4     & 20        & 0.90                & 4.694       & 5          & 40    & 55    \\ \hline
\end{tabular}
\end{table}

\begin{figure}[h!]
\centering
\hbox{\hspace{-1cm}\includegraphics[width=1.1\textwidth]{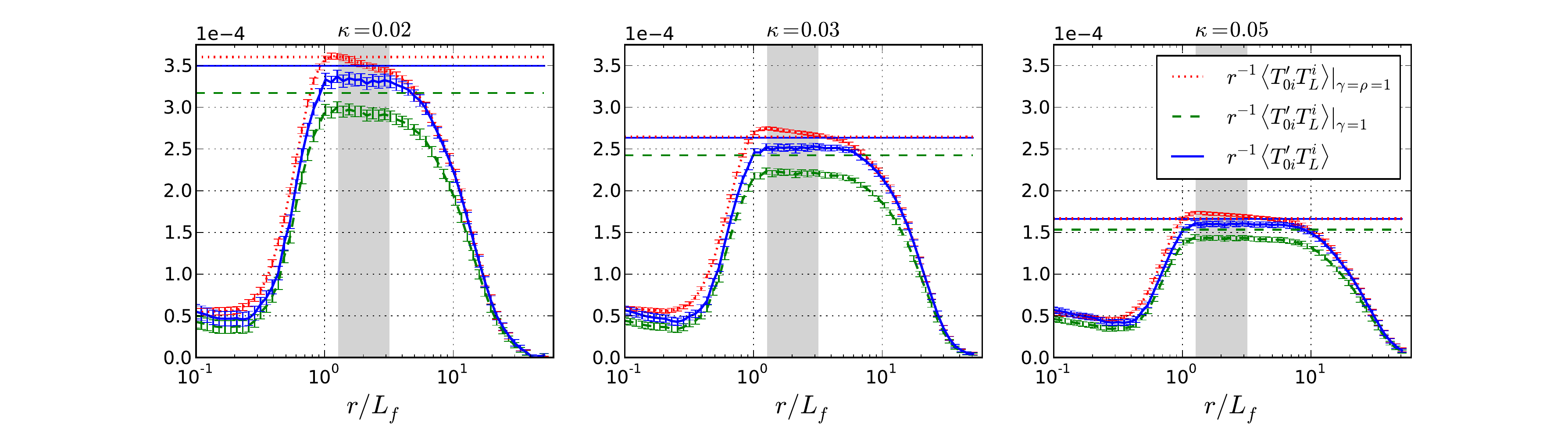}}
\caption{The relativistic correlation function $\la T_{0i}^\prime T^i_L \ra $ (solid blue) and its non-relativistic compressible and incompressible counterparts, $\la T_{0i}^\prime T^i_L \ra\vert_{\gamma=1} $ (dashed green) and $\la T_{0i}^\prime T^i_L \ra\vert_{\gamma=\rho=1} $ (dotted red), respectively, compensated by $r^{-1}$. From left to right: cases with dissipation strength $\kappa=0.02,\: 0.03, \: 0.05$, respectively. Each prediction for the inverse-cascade range $r/L_f \sim (10^0,10^1)$ is plotted as a horizontal line with matching line style. The predictions follow from Eq.~\eqref{eq:inversecascade_WS} and evaluations at $\gamma=1$ or $\gamma=\rho=1$ thereof. Note that the centre and right plots have a nearly indistinguishable prediction for the relativistic and incompressible correlation functions (solid blue and red dotted lines, respectively). Error bars correspond to the statistical uncertainty $\sigma/\sqrt{N}$ for each value of $r/L_f$, where $N$ is the sample size and $\sigma$ is the sample standard deviation. The shaded grey area indicates the range of $r/L_f$ over which we fit a power-law, and we use the same range across all cases to ensure a fair comparison. Note the linear vertical scale, which accentuates deviations from the expected power law.} \label{fig:corrs_inverse_cascade_compensated}
\end{figure}

In Fig.~\eqref{fig:corrs_inverse_cascade_compensated} we plot the relativistic correlation function appearing in Eq.~\eqref{eq:inversecascade_WS}, $\la T_{0i}^\prime T^i_L \ra$, compensated for the expected scaling $r^{-1}$, with a linear vertical scale to help distinguish different power laws. We also plot the incompressible limit of that correlation function, obtained by setting $\rho=\gamma=1$ (herein ``the incompressible correlation"), as well as a non-relativistic but compressible version obtained by setting $\gamma=1$ only (herein ``the compressible correlation"). The former is equivalent to known results from incompressible Navier-Stokes turbulence (see Sec.~\eqref{sec:inverseincompressiblelimit}), while the latter can be obtained from the left-hand side of Eq.~\eqref{eq:inversecascade_WS} using the non-relativistic perfect fluid energy-momentum tensor, which is just the relativistic one with $\gamma=1$. We use these comparisons to separately gauge the degree to which compressive and relativistic effects are important. In addition,  we also include the predictions for each case, in matching colour, obtained from Eq.~\eqref{eq:inversecascade_WS} and evaluations at $\gamma=1$ and $\gamma=\rho=1$ thereof. Error bars correspond to the statistical uncertainty $\sigma/\sqrt{N}$.

\begin{figure}[h!]
\centering
\hbox{\hspace{4.5cm}\includegraphics[width=0.5\textwidth]{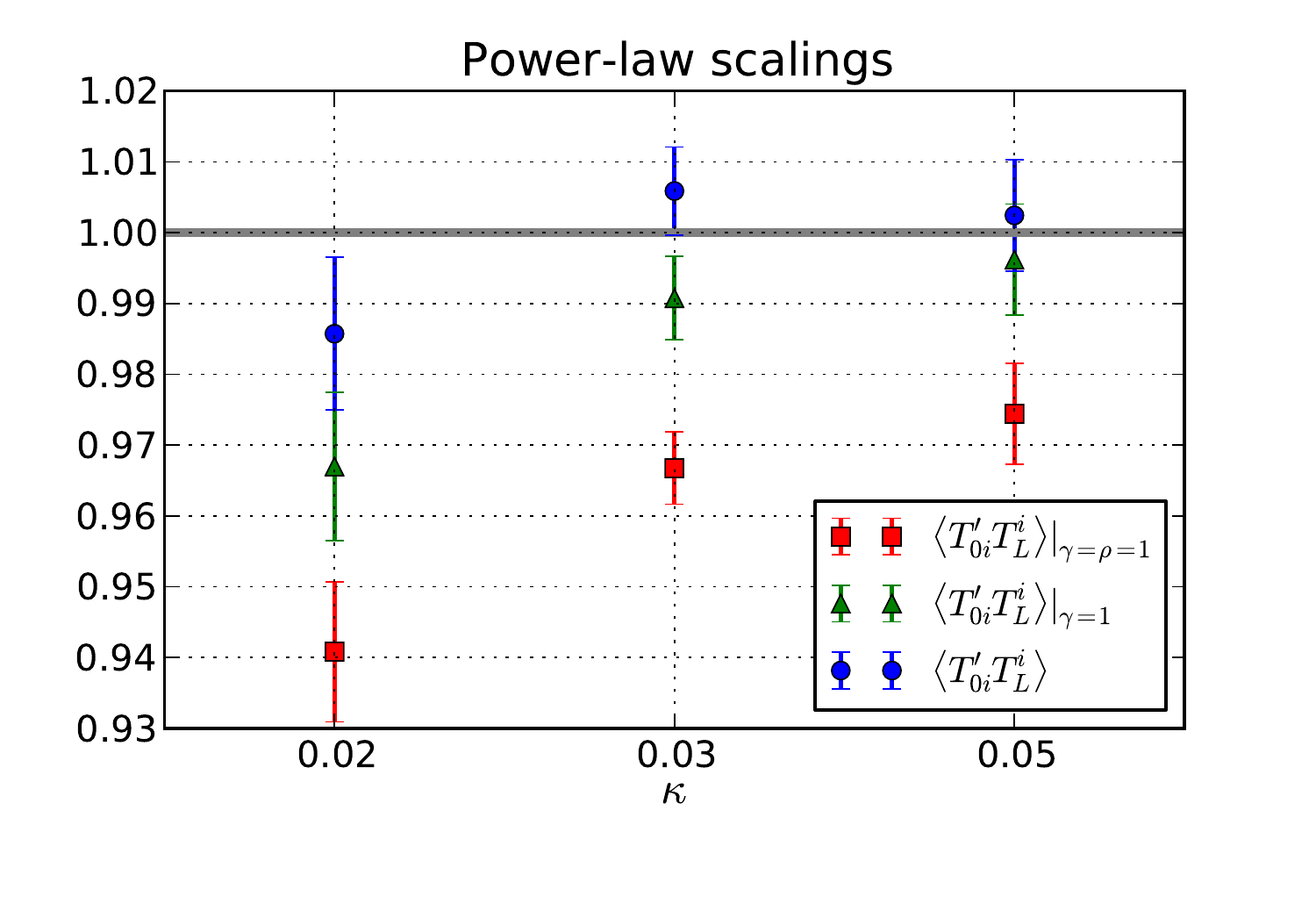}}
\caption{Least-squares power law fits for the relativistic correlation function $\la T_{0i}^\prime T^i_L \ra $ (circles) and its non-relativistic compressible and incompressible counterparts, $\la T_{0i}^\prime T^i_L \ra\vert_{\gamma=1} $ (triangles) and $\la T_{0i}^\prime T^i_L \ra\vert_{\gamma=\rho=1} $ (squares), respectively. All three dissipation cases are displayed. Error bars correspond to the standard deviation of the fitted power law scaling, obtained via random resampling with replacement ($10^3$ trials). The relativistic correlation function outperforms its compressible and incompressible counterparts across all cases, with a monotonic degradation observed for the latter two as the dissipation strength is decreased (and correspondingly, as the rms Mach number is increased). } \label{fig:powerlaw_fits_inverse_cascade}
\end{figure}

For ease of comparison across cases, each plot has the same vertical axis range. As it is clear from the figure, we observe a progressive degradation of the scaling of the incompressible and compressible correlation functions as dissipation is weakened (and thus Mach number grows), while the relativistic one predicted in~\cite{WS2015} outperforms. This is shown quantitatively in Fig.~\eqref{fig:powerlaw_fits_inverse_cascade}, where we display power law fits performed over the shaded interval of Fig.~\eqref{fig:corrs_inverse_cascade_compensated}. The shaded interval is the same across all cases in order to make a fair comparison, and is chosen to capture the power law observed in the $\kappa=0.02$ case (which has the narrowest scaling range). As dissipation is weakened, a monotonic shallowing of the best-fit power law is observed for both the incompressible and compressible correlation functions. This trend is more significant for the incompressible correlation function. The absolute performance of the relativistic correlation function is superior to the compressible and incompressible correlation functions across all cases, and its relative performance improves as dissipation is decreased (i.e. differences in best-fit power law become larger).

The relativistic correlation function, although exhibiting power-law scaling $\sim r$ in all cases, nonetheless exhibits an increasing disagreement with the \emph{magnitude} of the prediction in Eq.~\eqref{eq:inversecascade_WS} (see Fig.~\eqref{fig:corrs_inverse_cascade_compensated}). In the most extreme case ($\kappa =0.02$, $v_{rms}/c_s = 0.19$), the overall magnitude is less than the prediction by $\sim 4\%$. Our numerical ensemble of flows may be biased towards lower magnitudes, since runs with sufficiently large fluctuations from the random force can become numerically unstable and fail. As dissipation is weakened, this occurs more often. Thus, it is possible that the increasing disagreement of the magnitude of $\la T_{0i}^\prime T^i_L \ra$ and $(1/2)\epsilon r$ is an artefact of this bias. A high-resolution shock-capturing implementation could determine whether this increasing disagreement is a real effect.

%
%
\subsection{Direct-cascade simulation} \label{sec:directcascaderesults}
\begin{figure}[h!]
\centering
\hbox{\hspace{0cm}\includegraphics[width=\textwidth]{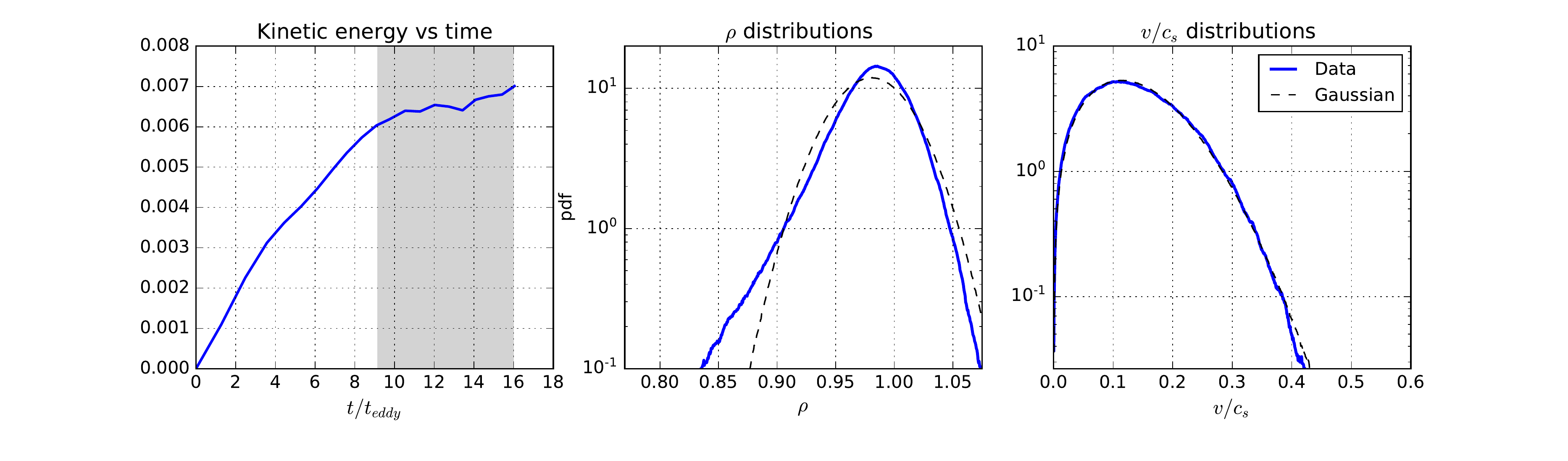}}
\caption{(Left): Average Newtonian specific kinetic energy plotted as a function of dimensionless time for the direct-cascade simulation. The eddy turnover time is defined as $t_{eddy} = 2\pi/\omega_{rms}$. The shaded region indicates the interval of time over which all other quantities are averaged. (Centre and Right): Probability distribution functions for the energy density $\rho$ (Centre, blue, solid) and the Mach number $v/c_s$ (Right, blue, solid), where $c_s = c/\sqrt{2}$, plotted on a semi-log scale. For comparison, purely Gaussian distributions are plotted (black, dashed) with the average and standard deviation matched to the data. The standard deviation of the $\rho$ and $v/c_s$ distributions are $(0.0336,0.0758)$, respectively. The rms Mach number is $0.11$. These properties indicate a weakly compressible flow.} \label{fig:pdfs_direct_cascade}
\end{figure}

To simulate the direct-cascade of a $(2+1)$-dimensional conformal fluid, we instead use an external force with support only around $k_f=7$, as described by Eq.~\eqref{eq:spikeprofile}. As in our inverse-cascade simulations, we use 4th-order numerical dissipation as in Sec.~\eqref{sec:dealiasing}, with the choice $\kappa = 0.01$. However, in contrast to our inverse-cascade simulations, here we use a large-scale dissipation mechanism known as 4th-order \emph{hypofriction}, which takes the form of a term $-\mu \nabla^{-4}S^{i}$ on the right-hand side of Eq.~\eqref{eq:Seom}. We compute the inverse Laplacians spectrally, setting constant modes to zero. Such a term has power restricted to large scales, and terminates the brief inverse cascade from $k=7$ towards $k=0$. We find the value $\mu=0.15$ to be adequate for preventing a build-up of energy (and eventual condensation) at large scales. An energy condensate would be characterized by continued energy growth and the emergence of two dominant vorticies of opposing parity superposed on a noisy flow (see eg.~\cite{Chertkov:2007}). Statistical quantities are averaged over the shaded interval of time indicated in Fig.~\eqref{fig:pdfs_direct_cascade} (Left), which consists of $N=10$ snapshots separated by $\delta T \sim 0.68$. The interval is chosen to maximize the number of snapshots available (to minimize statistical fluctuations), while remaining in a regime that roughly resembles a steady-state. Note that the measured correlations are not significantly affected if the temporal average begins slightly earlier or later. The average time step over this interval is $10^{-3} t_{eddy}$, where $t_{eddy}=2\pi/\omega_{rms} \sim 0.5$ is also averaged over the shaded interval. The injection rate of $(1/2)\la T_{0i}T_0^i\ra$ is measured initially to be $\epsilon = 2\times 10^{-4}$. For reference, the characteristic time as per $v_{rms}$ is $L/v_{rms} \sim 26$, where we take $L=2\pi/3$ since the maximum of the energy spectrum occurs at $k=3$. The rms Mach number over the shaded interval is $\sim 0.11$, once again indicating the weakly compressible regime.

\begin{figure}[h!]
\centering
\hbox{\hspace{0cm}\includegraphics[width=\textwidth]{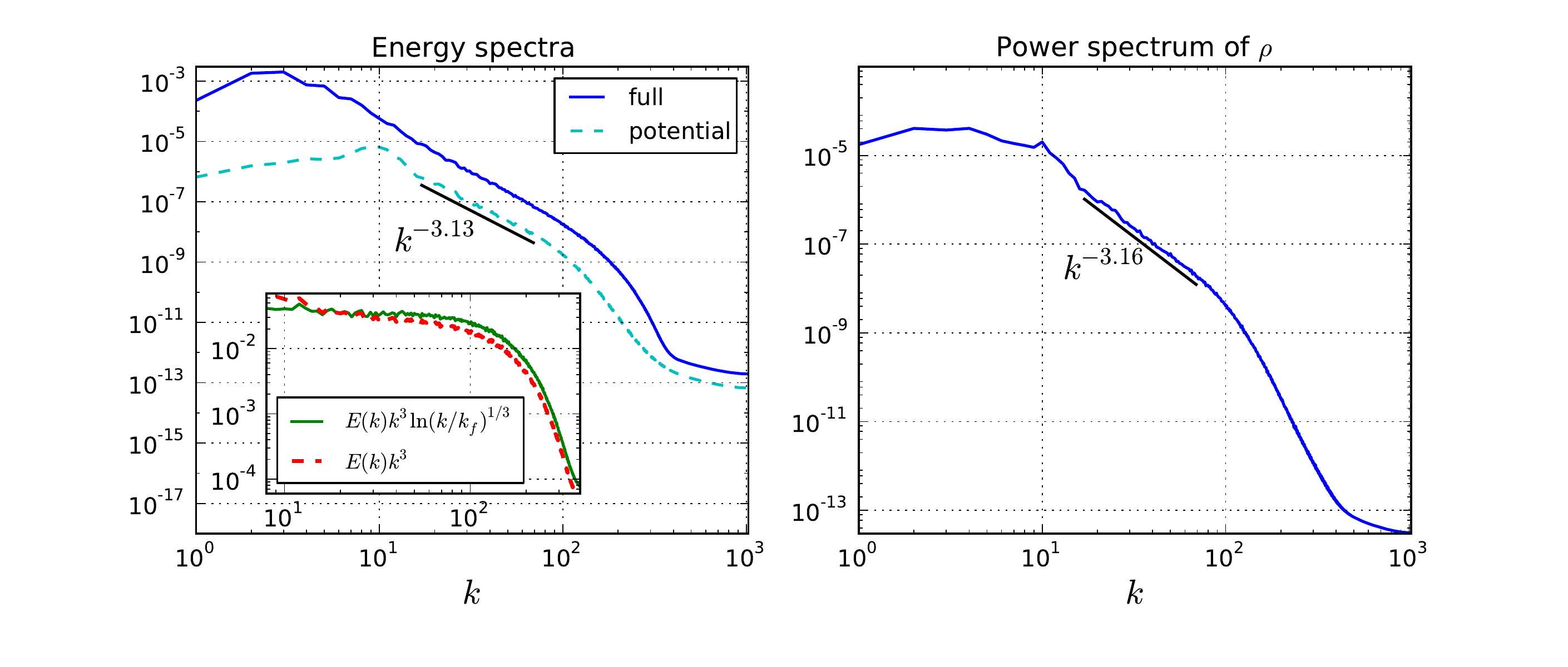}}
\caption{(Left): Newtonian energy spectra of the inverse-cascade simulations plotted on a log-log scale. The spectra corresponding to the full velocity field (solid, blue) and the curl-free potential part (dashed, cyan) are displayed. A least-squares best fit power law of the potential part $\sim k^{-3.13}$ over the range $k\in [17,70]$ is displayed. The inset displays the full spectrum compensated by $k^3 \ln{(k/k_f)}^{1/3}$ (solid, green) or $k^3$ only (dashed, red), showing that the logarithmic correction provides a better fit than the pure power law. (Right): The power spectrum of the energy density $\rho$, with the best fit power law of $k^{-3.16}$ over the range $k\in [17,70]$ displayed.} \label{fig:direct_cascade_all_spectra}
\end{figure}

In Fig.~\eqref{fig:pdfs_direct_cascade} (Left), we display the average Newtonian specific kinetic energy of the fluid as a function of time. As mentioned, we average various quantities over the shaded interval of time. The energy is beginning to plateau over this interval, however it continues to grow slowly. If evolved longer, the compressive component of the velocity begins to dominate over the curl-free part. To study such a regime more accurately, a Riemann solver would be desirable in order to more faithfully capture the dominant shockwave phenomena. Since we are instead using artificial high-order numerical dissipation, we choose to restrict our analysis to earlier times, when the compressive component of the velocity field is still subdominant ($\sim 10\%$ of the total energy at a given scale $k$ - see Fig.~\eqref{fig:direct_cascade_all_spectra} (Left)). Our high-order dissipation also results in large bottleneck effects at later times, which contaminate a rather large portion of the inertial range.

In Fig.~\eqref{fig:pdfs_direct_cascade} (Centre and Right), we display the probability distributions of the energy density (Centre, blue, solid) and Mach number (Right, blue, solid). For comparison, in both cases we also plot Gaussian distributions (black, dashed) with average and standard deviation matched to the data. Similar to the inverse-cascade simulations, the Gaussian provides a good fit to the Mach number pdf (although with weaker hints of a non-Gaussian tail in this case), whereas the energy density pdf exhibits a stronger, exponential tail towards smaller values.

In Fig.~\eqref{fig:direct_cascade_all_spectra} we display the power spectra of the velocity (Left) and energy density (Right). The full energy spectrum of the flow (blue, solid), together with the energy spectrum of the compressive, curl-free, potential part of the velocity (cyan, dashed). The latter is seen to be subdominant by a factor of $\sim 10$ over the range $k\in [10,100]$, which qualitatively corresponds to the direct-cascade interial range. The potential spectrum is fit by a power law $k^{-3.13}$ over this range, while for the full spectrum we observe $k^{-3}$ scaling with the multiplicative logarithmic correction $\ln(k/k_f)^{-1/3}$. The inset shows the full spectrum compensated by $k^{3}$ with and without the logarithmic correction, with the presence of the logarithmic correction being favoured (flatter curve). In the literature, the presence of this correction seems to depend on several factors, including the length of time over which the average is taken, and the presence of large-scale dissipation~\cite{Pasquero2002,Alvelius2000,Chen2003,Vallgren2011}. 

\begin{figure}[h!]
\centering
\hbox{\hspace{1.7cm}\includegraphics[width=0.8\textwidth]{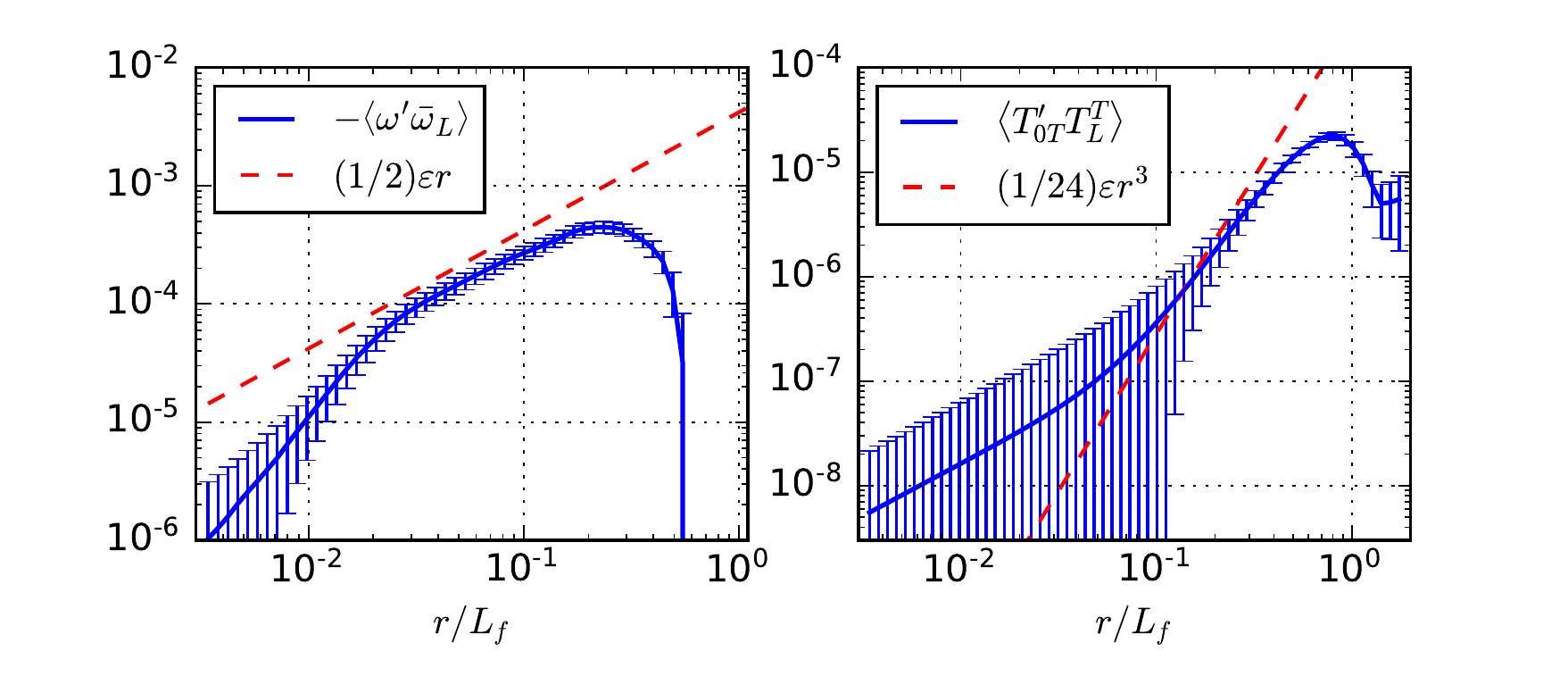}}
\caption{(Left): The relativistic correlation function $-\la \omega^\prime \bar{\omega}_L \ra$ (blue, solid) plotted with its prediction in the direct-cascade range, $(1/2)\varepsilon r$ (red, dashed), as per Eq.~\eqref{eq:directcascade_WS}. The correlation function $-\la \omega^\prime \bar{\omega}_L \ra$ is fit by $\sim r^{2.24}$ at shorter distances and $\sim r^{0.840}$ at larger distances, with the transition occurring near $r/L_f \sim 2\times 10^{-2}$. (Right): The relativistic correlation function $\la T^\prime_{0T} T^T_L \ra$ (blue, solid) plotted with its prediction in the direct-cascade range, $(1/24)\varepsilon r^3$ (red, dashed), as per Eq.~\eqref{eq:directcascadeWS_weird}. The correlation function $\la T^\prime_{0T} T^T_L \ra$ is fit by $\sim r^{1.05}$ at shorter distances and $\sim r^{2.40}$ at longer distances, with the transition occurring near $r/L_f \sim 10^{-1}$. Error bars correspond to the statistical uncertainty $\sigma/\sqrt{N}$ for each value of $r/L_f$, where $N$ is the sample size and $\sigma$ is the sample standard deviation. Note that the statistical uncertainty at short distances is sufficiently large that the sign of the correlation functions is uncertain there. The corresponding compressible and incompressible limits of these correlation functions (obtained by setting $\gamma=1$ or $\gamma=\rho=1$, as in Sec.~\eqref{sec:inversecascaderesults}) do not behave significantly differently. This allows us to approximate $\varepsilon$ by its incompressible limit $((1+w)/w)^2 \rho^2 \epsilon_\omega$ as in Sec.~\eqref{sec:direct-cascade-limit}, and we take $\epsilon_\omega$ to be the initial enstrophy growth rate (before dissipation mechanisms become important). However, we substitute $\la \rho^2 \ra \sim 0.96$ in place of $\rho^2$, which slightly improves agreement with the predictions.} \label{fig:corrs_direct_cascade}
\end{figure}

In Fig.~\eqref{fig:corrs_direct_cascade}, we display the two measured correlation functions for which we have predictions in the direct cascade (Eqs.~\eqref{eq:directcascade_WS} and~\eqref{eq:directcascadeWS_weird}). Errors again correspond to the statistical uncertainty $\sigma/\sqrt{N}$. We find reasonable agreement in the case of Eq.~\eqref{eq:directcascade_WS} (Left), and less so in the case of Eq.~\eqref{eq:directcascadeWS_weird} (Right). The measured power laws are $r^{0.84}$ (Left) and $r^{2.4}$, as compared to the predictions of $r$ and $r^3$, respectively. However, we note that the non-trivial factors of $1/2$ (Left) and $1/24$ (Right) yield marked agreement in magnitude. We suspect that by decreasing contamination from our modified large- and small-scale dissipation mechanisms, agreement with our predictions would improve, since in our inverse-cascade simulations we found that removing large-scale dissipation altogether improved agreement with our predictions significantly. We also note that the statistical uncertainty at short length scales is large enough that the sign of the correlation functions is uncertain there. We estimate the quantity $\varepsilon$ by its incompressible limit $((1+w)/w)^2 \rho^2 \epsilon_\omega$, with a further substitution of $\la \rho^2 \ra = 0.96$ in place of $\rho^2$ (which improves agreement slightly). This estimate is justified by the fact that the correlation functions we measure do not differ significantly from their incompressible counterparts (i.e. setting $\rho=\gamma=1$) in this regime.

%
%
\section{Discussion}\label{sec:discussion}
\begin{figure}[h!]
\centering
\hbox{\hspace{3.5cm}\includegraphics[width=0.6\textwidth]{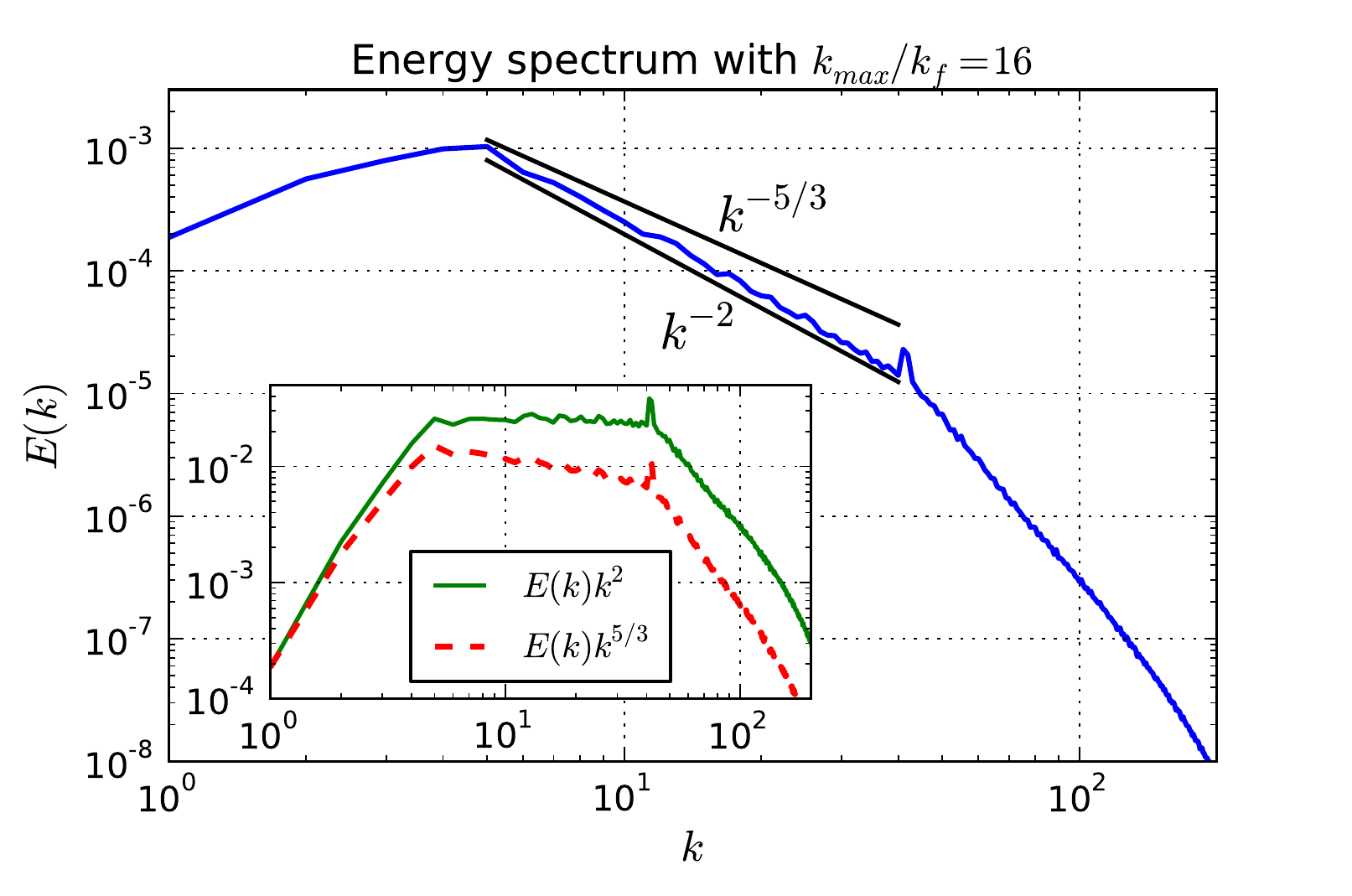}}
\caption{Energy spectrum $E(k)$ (blue, solid) with forcing active at $k_f=42$, such that $k_{max}/k_f = 16$, where $k_{max} = N/3$ and $N=2048$. Power laws $k^{-2}$ and $k^{-5/3}$ (black, solid) are shown for comparison. The inset displays compensated spectra $E(k)k^2$ (green, solid) and $E(k)k^{5/3}$ (red, dashed). The spectrum is well-represented by $k^{-2}$, rather than $k^{-5/3}$, consistent with~\cite{Scott:2007} (albeit for a conformal fluid in our case). The best fit power-law slope over the range $k\in [5,40]$ is $-2.047$.} \label{fig:k42_spectrum}
\end{figure}

As mentioned in Sec.~\eqref{sec:inversecascaderesults}, we observed that our energy spectra approach a $k^{-2}$ scaling in the inverse-cascade range. We point out that it was observed in~\cite{Scott:2007} that the incompressible case exhibits the same scaling provided large-scale friction is absent and the direct cascade is sufficiently resolved ($k_{\text{max}}/k_f \geq 16$, where we define $k_{\text{max}}=\mathcal{N}/3$). It thus becomes a prescient question whether a sufficiently resolved direct cascade in the conformal fluid case will yield the same result. To answer this, we perform an ensemble of $20$ simulations with $\mathcal{N}=2048$ and $k_{\text{max}}/k_f = 16$, with a forcing profile given by Eq.~\eqref{eq:spikeprofile}. The resulting energy spectrum is displayed in Fig.~\eqref{fig:k42_spectrum}, with the inset displaying the same spectrum compensated by either $k^{2}$ or $k^{5/3}$. After filtering out the modes $k \in [0,5]$ (i.e. all modes less than the maximum of the spectrum), the rms Mach number for this flow is $\sim 0.11$. We perform this filtering so as to have a more fair comparison of rms Mach number with our inverse-cascade simulations in Sec.~\eqref{sec:inversecascaderesults}, which we remind were $\sim 0.14$, $0.17$, and $0.19$. As evident from Fig.~\eqref{fig:k42_spectrum}, the spectrum clearly favours a $k^{-2}$ description in the inverse-cascade range, and not the Kolmogorov/Kraichnan $k^{-5/3}$ description. The best-fit power law over the range $k \in [5,40]$ yields $k^{-2.047}$. 

Interestingly, we note in passing that this $k^{-2}$ scaling of the energy spectrum would change the result of the purported calculation of the fractal dimension of a turbulent $(3+1)$-dimensional AdS-black brane presented in~\cite{Adams:2013vsa} to $D=3$ (rather than $D=3+1/3$). An analysis of this will be reported elsewhere~\cite{fractalhorizon}.

We also point out that, despite the narrow inverse-cascade range, we nonetheless observe a similarly narrow $\sim r^{0.95}$ power law scaling in the same correlation functions analyzed in Sec.~\eqref{sec:inversecascaderesults} (not shown). This suggests that the scaling relation Eq.~\eqref{eq:inversecascade_WS} continues to hold with a more resolved direct-cascade range.

\begin{figure}[h!]
\centering
\hbox{\hspace{0cm}\includegraphics[width=\textwidth]{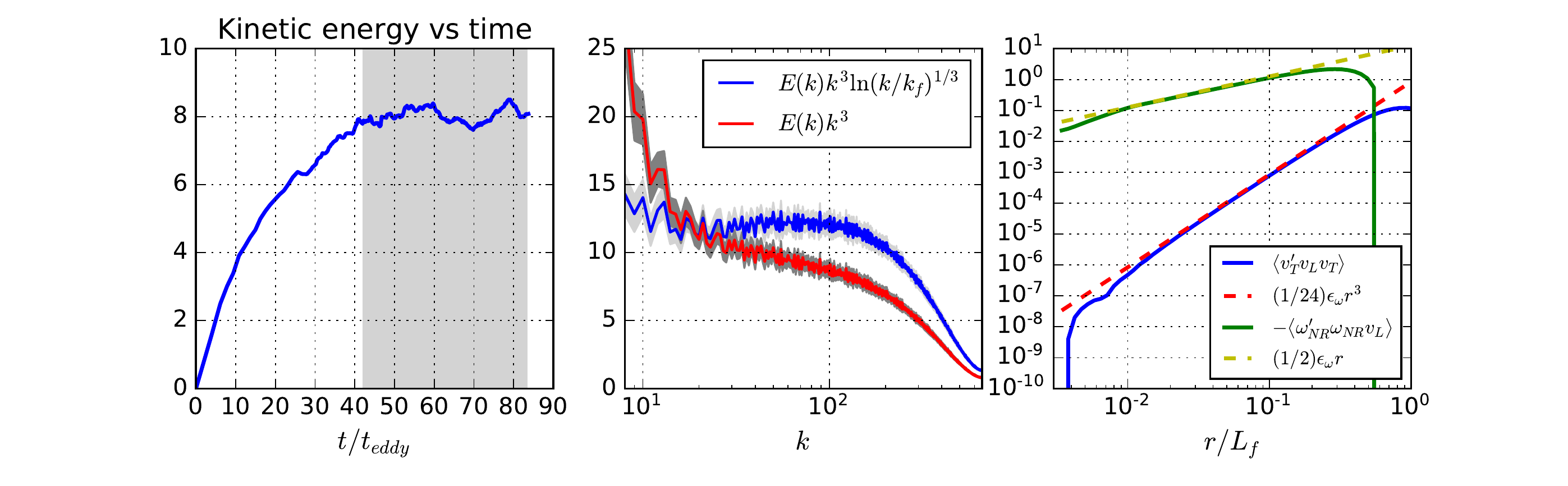}}
\caption{Data from our simulation of an incompressible Navier-Stokes fluid. (Left): The average specific Newtonian kinetic energy plotted as a function of time. The shaded interval corresponds to the interval over which we compute statistical averages. (Centre): The energy spectrum $E(k)$ compensated by either $k^3 \ln{k/k_f}^{1/3}$ or $k^3$. The shaded envelopes correspond to $5\times$ the statistical error $\sigma/\sqrt{N}$. The logarithmic correction is evidently favoured. (Right): The correlation functions (solid) and their predictions (dashed) corresponding to the incompressible counterparts of Eqs.~\eqref{eq:directcascade_WS} and~\eqref{eq:directcascadeWS_weird}.} \label{fig:NSstuff}
\end{figure}

In Sec.~\eqref{sec:directcascaderesults}, we observed hints of the predicted scaling of the correlation functions displayed in Fig.~\eqref{fig:corrs_direct_cascade}. By instead simulating an incompressible fluid (as per Sec.~\eqref{sec:implementation-NS}), for which shockwave phenomena are not present, we can measure the incompressible limits of Eqs.~\eqref{eq:directcascadeWS_limit} and~\eqref{eq:directcascadeWS_weird_limit} with greater statistical significance. Conformal fluids have been shown to possess a scaling limit to an incompressible Navier-Stokes fluid~\cite{Bhattacharyya2009,Fouxon2008conformal}. We present the results of our simulation in Fig.~\eqref{fig:NSstuff} (Right). The enstrophy injection rate is set a priori to $\epsilon_\omega = 28$. We use 4th-order hypofriction $-\lambda \nabla^{-4} \omega$ and hyperviscosity $-\nu_4 \nabla^4 \omega$, with $\lambda=0.15$ and $\nu_4 = 10^{-10}$. The forcing profile is given by 
\begin{eqnarray}
\la f_\omega f^\prime_\omega \ra \propto {\begin{cases}
 1 \:\:\:\: k \sim k_f \\
 0 \:\:\:\: \text{otherwise} \end{cases}},
\end{eqnarray}
and the injection scale is set to $k_f = 7$. The average specific kinetic energy is plotted as a function of time in Fig.~\eqref{fig:NSstuff} (Left), with the averaging interval shaded gray. The energy spectrum compensated by $k^3$ and $k^3\ln{(k/k_f)}^{1/3}$ is displayed in Fig.~\eqref{fig:NSstuff} (Centre), with the shaded envelopes indicating $5\times$ the statistical uncertainty $\sigma/\sqrt{N}$. The logarithmic correction is clearly favoured. In Fig.~\eqref{fig:NSstuff} (Right), we plot the correlation functions $\la v_T^\prime v_L v_T \ra$ and $-\la \omega^\prime_{\text{NR}} \omega_{\text{NR}} v_L \ra$, together with their respective predictions in the direct cascade $(1/24)\epsilon_\omega r^3$ and $(1/2)\epsilon_\omega r$. The agreement is very much improved over Fig.~\eqref{fig:corrs_direct_cascade}, which suggests that despite the low Mach number in our direct cascade simulation of the conformal fluid, that situation is nonetheless quite different from the incompressible case (at least insofar as the numerical challenges are greater in the former case, eg. small scales being contaminated by bottleneck effects).

\begin{figure}[h!]
\centering
\hbox{\hspace{1cm}\includegraphics[width=0.9\textwidth]{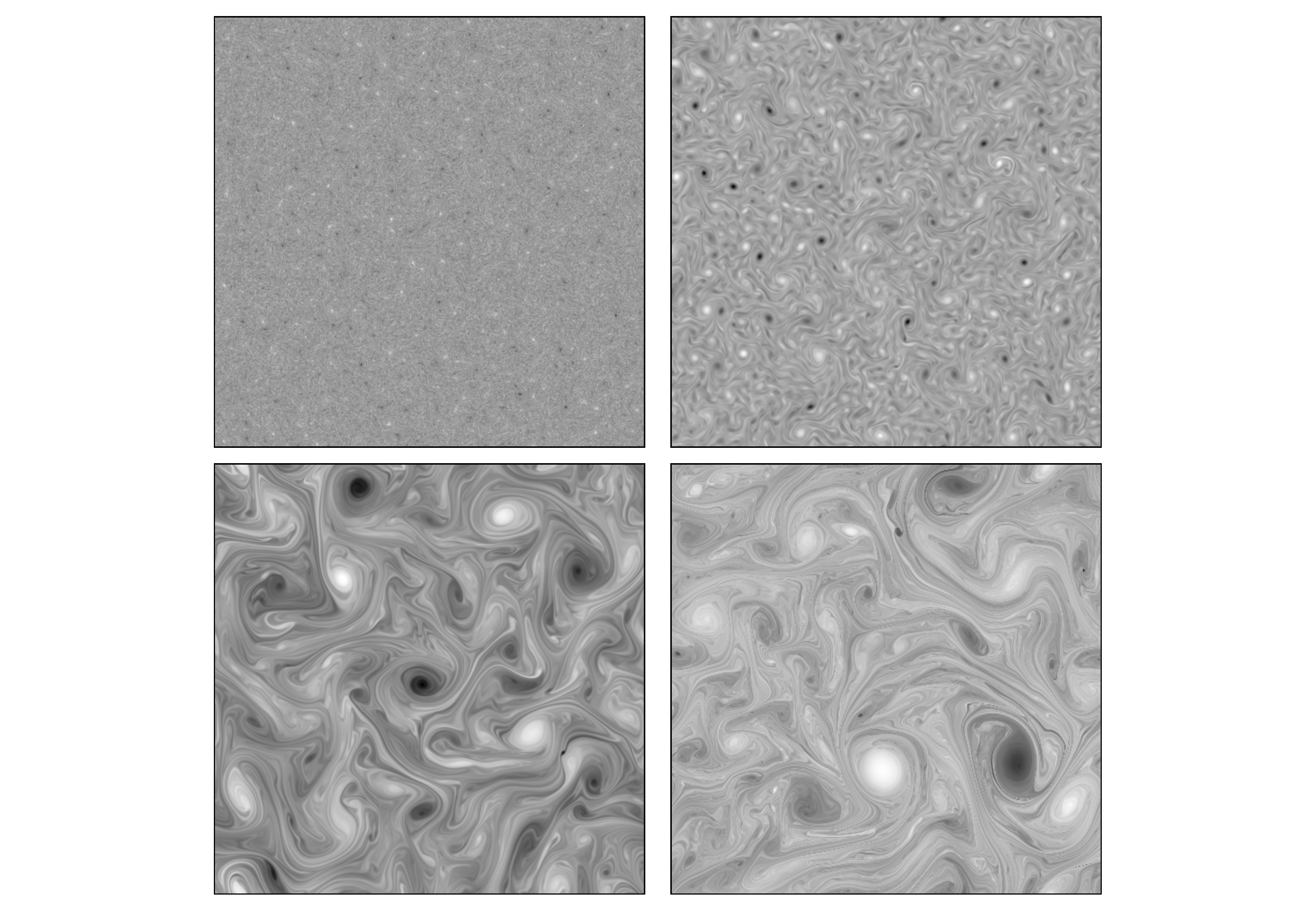}}
\caption{Snapshots of vorticity. (Top Left): Conformal fluid inverse-cascade with $k_f=203$ ($\kappa=0.02$ case). (Top Right): Conformal fluid inverse cascade with $k_f=42$. (Bottom Left): Conformal fluid direct cascade with $k_f=7$. (Bottom Right): Incompressible fluid direct cascade with $k_f=7$.} \label{fig:allvorts}
\end{figure}

Finally, in Fig.~\eqref{fig:allvorts} we display snapshots of the vorticity from several of our simulations. By doing so, we intend to provide intuition as to how the inverse and direct cascades appear in real space. In particular, the stretching and mixing of vorticity isolines characteristic of the direct-cascade range are readily identified as `turbulence' qualitatively, where coherent features are seen over a variety of scales (see Fig.~\eqref{fig:allvorts} (Bottom Left) and (Bottom Right)). By contrast, the inverse-cascade range has a much noisier appearance, as in Fig.~\eqref{fig:allvorts} (Top Left). We have not observed an explicit acknowledgement of this qualitative fact in the literature, since numerical studies of the inverse-cascade range seldom include plots of the vorticity (eg.~\cite{Boffetta:2000}). Unless the direct-cascade range is resolved, the turbulent flow qualitatively appears as random noise -- but even if it is resolved, a clear hierarchy of scales is not apparent in the inverse-cascade range. In Fig.~\eqref{fig:allvorts} (Top Right) we show a mixed case with the forcing acting at an intermediate scale $k_f=42$. This case is a simulation targeting the inverse-cascade range, but the direct-cascade range is just beginning to be resolved as well. Consequently, vorticity isoline mixing is beginning to be apparent, superposed on top of a more noisy structure.

%
%
\acknowledgements
We would like to thank Michael Waite for detailed discussions on dealiasing and other matters. We also thank Guido Boffetta for sharing unpublished statistical values from the study~\cite{Boffetta:2000}, and Gregory Falkovich for an interesting discussion. This work
was supported by NSERC through a Discovery Grant and by CIFAR (L.L.). J.R.W.S. acknowledges support from OGS. 
This research was enabled in part by support provided by scinet (www.scinethpc.ca) and Compute Canada (www.computecanada.ca).
Research at Perimeter Institute is supported through Industry Canada and by the Province of Ontario through the Ministry of Research \& Innovation.

\bibliography{fluidbib}

\end{document}